# Topologically tunable polaritons based on two-dimensional crystals in a photonic lattice

L. Lackner[1], O.A. Egorov[2], A. Ernzerhof[1], C. Bennenhei[1], V.N. Mitryakhin[1], G. Leibeling[3], F. Eilenberger[3,4], S. Tongay[5], U. Peschel[2], M. Esmann[1] and C. Schneider[1,†]

[1]Institut für Physik, Fakultät V, Carl von Ossietzky Universität Oldenburg, 26129 Oldenburg, Germany.

[2]Institute of Condensed Matter Theory and Solid State Optics, Friedrich Schiller University, Jena, Germany

[3]Fraunhofer-Institute for Applied Optics and Precision Engineering IOF, 07745 Jena, Germany.

[4]Institute of Applied Physics, Abbe Center of Photonics, Friedrich Schiller University, 07745 Jena, Germany

[5]School for Engineering of Matter, Transport, and Energy, Arizona State University, Tempe, Arizona 85287, USA

†Corresponding author. Email: Christian.Schneider@uni-oldenburg.de

**Topological photonics is an emergent research discipline which interlinks fundamental aspects of photonics, information processing and solid-state physics. Exciton-polaritons are a specifically interesting platform to study topological phenomena, since the coherent light matter coupling enables new degrees of freedom such as tunable non-linearities, chiralities and dissipation. Room-temperature operation of such exciton-polaritons relies on materials comprising both, large exciton binding energies and oscillator strength. We harness widely spectrally tunable, room temperature exciton-polaritons based on a $WS_2$ monolayer in an open optical cavity to realize a polariton potential landscape which emulates the Su-Schrieffer-Heeger (SSH) Hamiltonian. It comprises a domain boundary hosting a topological, exponentially localized mode at the interface between two lattices characterized by different Zak-phases which features a spectral tunability over a range as large as 80 meV. Moreover, we utilize the unique tilt-tunability of our implementation, to transform the SSH-lattice into a Stark-ladder. This transformation couples the topologically protected defect mode to propagating lattice modes, and effectively changes the symmetry of the system. Furthermore, it allows us to directly quantify the Zak-phase difference $\Delta_{Zak} = (1.13 \pm 0.11)\pi$ between the two topological phases. Our work comprises an important step towards in-situ tuning topological lattices to control and guide light on non-linear chips.**

With the emergence of atomically thin crystals and the observation of giant light-matter coupling of excitons hosted in transition metal dichalcogenide (TMD) monolayers[1], the research field of exciton-polariton physics has experienced a paradigmatic shift: Excitons in TMD monolayers present a very large oscillator strength[2,3] and their large binding energies routinely facilitate the observation of exciton-polaritons from cryogenic up to room temperature[4–7]. Noteworthily, the sizeable polariton interaction[8,9] as well as polaritonic relaxations can be engineered in van-der-Waals heterostructures[10] with unprecedented degrees of freedom. Those assets, very recently, led to the first observations of polariton condensation in TMD microcavity architectures[11,12].

Beyond the study of fundamental aspects of light-matter coupling and in non-linear photonics, exciton-polaritons are considered one of the prime experimental platforms in the field of topological photonics. There, both the photonic[13] as well as the excitonic[14] part of the polaritonic mode can be actively utilized to manipulate the overall system topology, yielding significant observations of topological polariton laser oscillation[15,16], spin-momentum locking[17,18] and switching[19,20]. However, since these experiments were primarily carried out in monolithic, microstructured samples, reconfigurable tuning of these topological modes, and substantial modification of the emulated potentials has remained out of experimental reach.

In this work, we address this shortcoming, and implement a room temperature polariton systems, which is topologically nontrivial, spectrally tunable over an arbitrarily large range, and which can even be re-configured in terms of the symmetry of the Hamiltonian allowing us to measure its topological invariants, namely the Zak phase.

**Sample and experimental setup**

Our sample structure is schematically depicted in Fig. 1(a). It is composed of two distributed Bragg reflectors (DBRs) of $SiO_2/TiO_2$ (details can be found in the methods section), which are attached to XYZ nano-positioners with a goniometer and are separated by an air-gap. Thus, they form an 'open cavity' with full tunability of the air-gap length, relative mirror position and tilt. The atomically thin $WS_2$ layer is transferred onto the bottom DBR (see Fig. 1(c)) via scotch tape exfoliation followed by the dry-gel stamping method[21].

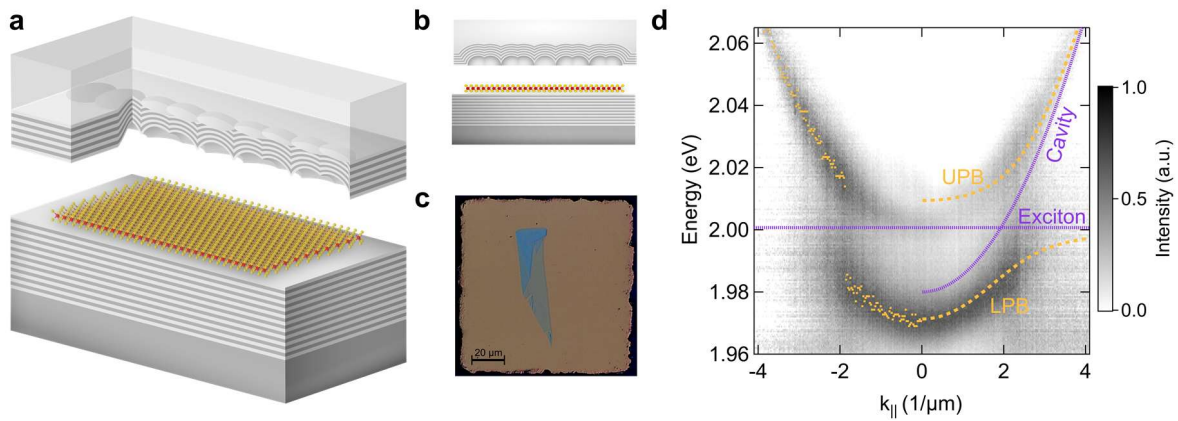

Fig. 1. **Open cavity and TMD monolayer sketch and strong coupling conditions. a** Sketch of the open cavity device, composed of two DBRs surrounding an air-gap. The bottom DBR sits on a mesa of 100 x 100 µm². The $WS_2$ monolayer is placed on top of the bottom DBR, in the maximum of the cavity field. The top DBR has a 1D SSH chain of spherical cap indentations implemented. Both DBRs can be independently displaced, ensuring full control over spatial and spectral tuning. **b** Side view of the open cavity device sketch to highlight the spherical cap indentations and the central domain boundary. **c** Microscope image of the transferred $WS_2$ monolayer on top of the DBR mesa. **d** Dispersion relation of the polariton emission for a $L_{cav}$ ~ 1.88 µm, under weak excitation with a non-resonant continuous wave laser. For negative momentum values ($k_\parallel$<0) the peak location, determined by peak fitting of the polariton modes in the dispersion relation, is indicated by orange dots. As a guide to the eye the excitonic and photonic (upper and lower polariton) modes are represented with violet (orange) lines for positive momentum values ($k_\parallel$>0).

The strong coupling regime between excitons in the $WS_2$ monolayer and the photonic resonance of our cavity is confirmed via momentum resolved photoluminescence spectroscopy (see methods section for details on the setup design). In Fig. 1(d), we depict an exemplary dispersion relation from our device, which was recorded using non-resonant optical pumping (532 nm wavelength, pump power of $P = 4.7\ \mu W$, and a Gaussian spot size with a FWHM diameter of ~3.6 µm). The length of the air-gap in the cavity was tuned to a distance of ~1.88 µm. The spectrum is characterized by the two dispersive polariton branches, separated by a Rabi-splitting of 32 meV. The spectrum matches the dispersion relation of a coupled oscillator (see Supplementary Note 2) with a longitudinal resonance mode of the open cavity ($E_C = 1.98\ eV$) and the exciton energy of the WS$_2$ monolayer ($E_X = 2.001\ eV$), yielding a negative detuning of $\Delta = E_C - E_X = -21\ meV$. As we will show later in this manuscript, this detuning can be changed at will via re-positioning of the top DBR.

As the next step, the confinement of TMD-polaritons is engineered via focused Gallium ion beam (FIB) milling to mechanically shape spherical cap traps of variable diameter, depth and arrangement in the glass carrier which supports the top DBR prior to the deposition of the mirror. In this work, we restrict ourselves to a one-dimensional chain of spherical cap indentations[22], which are geometrically arranged in a staggered configuration (see Fig. 1(b)). Since the inter-trap distance dictates the photonic hopping strength, each homogeneous domain of the lattice emulates a potential which is represented by the following expression[23]:

$$H = v \sum_{i=1}^{N} (|a_i\rangle\langle b_i| + h.c.) + w \sum_{i=1}^{N-1} (|a_{i+1}\rangle\langle b_i| + h.c.) \quad (1)$$

and which is commonly referred to as the Su-Schrieffer-Heeger model. Here, $v$ and $w$ denote the intra- and intercell hopping amplitudes, respectively. We denote the two sites in the *i*-th unit-cell as $a_i$ and $b_i$. If translational invariance is maintained, the momentum space Hamiltonian can be expressed as

$$H(k) = \begin{bmatrix} 0 & v + we^{-ik} \\ v + we^{+ik} & 0 \end{bmatrix} \quad (2)$$

Diagonalization of this Hamiltonian yields a dispersion relation of Bloch modes, which form two bands (depicted by "+" and "-") separated by a band gap:

$$\Omega^{\pm}(k) = \pm\sqrt{v^2 + w^2 + 2vw \cos k} \quad (3)$$

This lattice is one of the most widely studied potentials in the field of topology due to its structural simplicity: Depending on the unit cell definition (respectively the sign of $v - w \neq 0$) its topological invariant, the Zak-phase, is:

$$\gamma_{Zak}^{\pm} = \int_{-\pi}^{\pi} i\langle u_k|\partial_k u_k\rangle dk \quad (4)$$

The integration runs over the full first Brillouin zone and $u_k$ is the cell-periodic part of the underlying Bloch modes. This expression becomes 0 for $v > w$ (topologically trivial) and $\pm\pi$ for $v < w$ (topologically nontrivial).

In the system introduced above, we connect two lattices with Zak-phases 0 and $\pm\pi$ at a domain boundary, which is expected to give rise to a zero energy mode in the mode spectrum pinned at the center of the gap of width $2|v - w|$. In real space, the mode is localized at the boundary. Its envelope decays exponentially to either side, and it has amplitude contributions on only one sub-lattice, i.e. it is sub-lattice polarized. This mode with zero energy is topologically protected from any perturbation in the lattice that does not break chirality of the Hamiltonian.

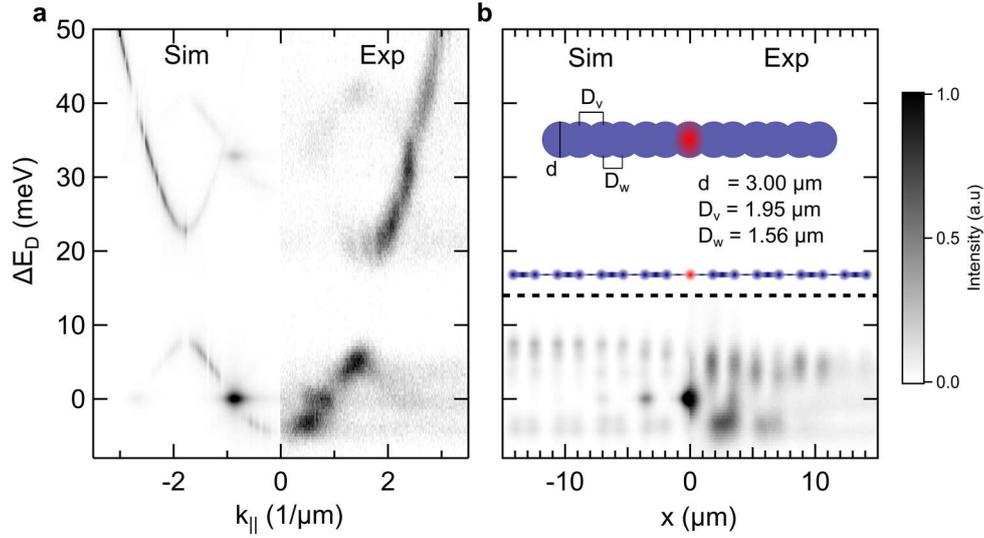

Fig. 2. **Real and Momentum Space properties of SSH-lattices a** Theoretical dispersion relation showing s- and p-bands with topologically nontrivial gaps in the s-band hosting a localized defect mode (left side, $k_\parallel<0$). Experimental dispersion relation (right side, $k_\parallel>0$). Energies have been offset by the energy of the defect mode in the s-band. **b** Top: Schematic drawing of the spherical cap moulds of diameter d and staggered center-to-center distance $D_{v,w}$. The defect state is located at the center of the chain and the excitation spot is indicated by the red circle. Bottom: Spatially resolved spectra of the SSH chain and the localized defect mode in the s-band; (x<0): Theoretical calculations, (x>0): Experimental emission spectrum.

Fig. 2 compares the calculated and measured dispersion relation (a), as well as the spectrally resolved real space distribution (b) of the main features in our lattice structure (details on the technique can be found in the Supplementary Note 3). Both, the experimental as well as the calculated spectra are plotted with respect to the energy difference $\Delta E_D$ to the topological defect state in the s-band. The simulated as well as the experimentally extracted dispersion relation features a significant spectral gap in the respective bands composed of coupled s-type orbitals. The size of this gap is directly correlated with the strength of the intra- and intercell hopping parameter $v$ and $w$. More importantly, a single state is found inside the gap. In real space this mode is exponentially localized in the s-band of the lattice and decays with a characteristic sub-lattice polarization[23], which we will discuss later in this manuscript in greater detail.

## Tuning topological lattices

The vertical displacement of the micro-structured top DBR changes the optical resonance conditions in our system. Consequently, we can control the eigen-energies of our optical lattice, as well as the energy of the topological mode in the SSH-gap on demand.

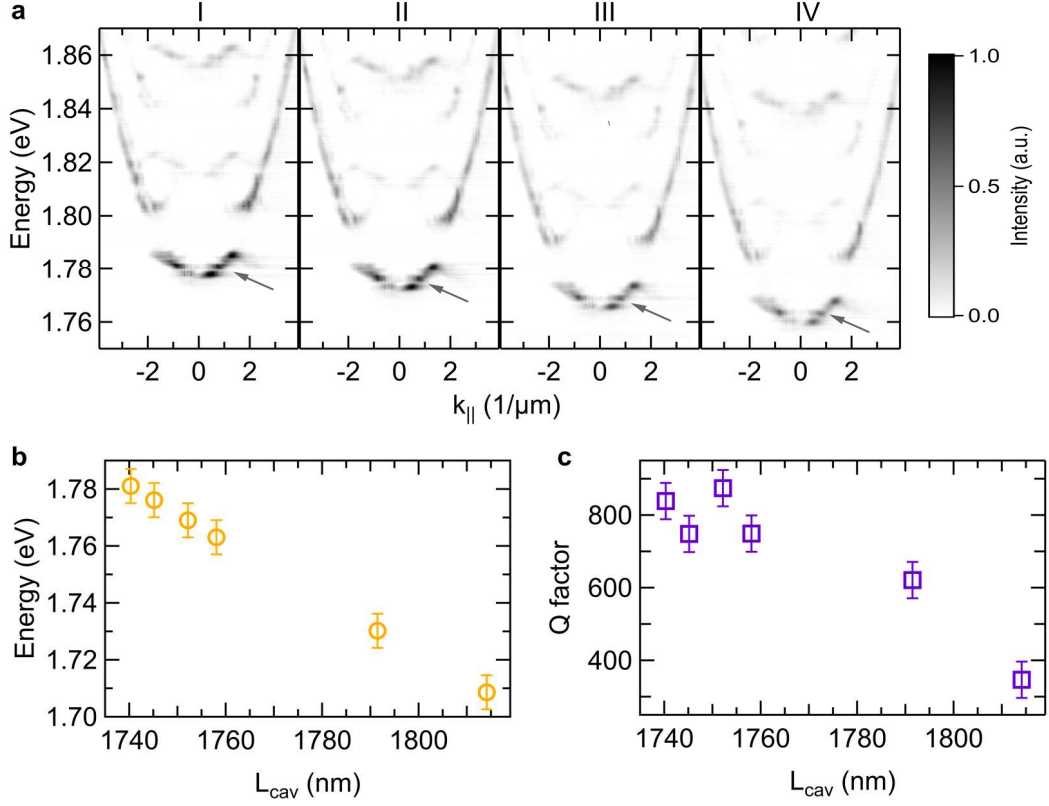

Fig. 3. **Spectrally widely tunable photonic topological lattices a** Momentum-resolved dispersion relations for varying cavity length. Tuning over 80 nm results in a 70 meV global energy shift of the dispersion relation, only limited by the spectral gain profile of the active TMD material. **b** Extracted spectral position of the localized SSH defect mode (arrows in (a)) in the s-band. **c** Experimentally determined Q-factor of the antibinding s-mode versus detuning.

This is clearly reflected in Fig. 3(a), where we plot a series of dispersion relations recorded for different separations of the two DBRs. The entire dispersion relation, including the topological mode, can be progressively tuned over an extraordinary range approaching 80 meV. We would like to point out that the tuning range is not limited by our apparatus, but merely by the spectral gain provided by the TMD monolayer which is integrated in the open cavity. In Fig. 3(b) we depict the energy of the topological mode of our polariton system versus the estimated cavity length. The corresponding Q-factor of the mode is plotted in Fi. 3(c). For cavity lengths shorter than 1780 nm we observe a saturation of the Q-factor around 800 which we explain by the mechanical contact of the two individual mirrors yielding a reduced dynamical broadening of the polariton mode due to the suppression of mirror vibrations.

This spectral tuning is essential to build versatile topologically nontrivial photonic devices. In the following, we will focus on the additional ability to modify the tilt between the two mirrors along the lattice via a built-in goniometer. Since the tilt angle between the mirrors is equivalent to continuously changing the cavity resonance and mode energy in real space, it corresponds to a modification of the real-space Hamiltonian of the system as:

$$H' = H + \mu \sum_{i=1}^{N} D \cdot (i-1)|a_i\rangle\langle a_i| + (iD - D_w)|b_i\rangle\langle b_i| \quad (5)$$

with $\mu$ the potential gradient, $D = D_w + D_v$ the unit cell length and $D_w$ the separation between two neighboring sites belonging to different unit cells $i$.

Since translational invariance is broken, this modification has a tremendous impact on the Hamiltonian, and its most fundamental properties are altered. As a first measurable quantity, we inspect the sub-lattice polarization of the mode in the topological gap of the s-band. We define a pseudo-spin polarization (PSP) for the $i$-th unit cell as

$$PSP_i = \left|\frac{I(b_i)-I(a_i)}{I(b_i)+I(a_i)}\right|, \qquad (6)$$

with $I$ the intensity of the emission on a given lattice site. The PSP is related to the total lattice polarization and has thus a strong connection with topological invariants as we will demonstrate below. It has also been measured in waveguide arrays by monitoring the mean displacement of an evolving field distribution.[24]

As we induce a potential gradient in the system, we also encounter a striking modification of this PSP. Fig. 4(a) shows a series of spatial cross-sections extracted at the energy of the defect mode in the s-band when systematically sweeping the potential gradient from negative to positive values. At zero gradient (see Fig. 4(b) for the corresponding, spectrally resolved image), we find according to equation (6) a $PSP_1$ of 0.93 for the first unit cell of the nontrivial domain (see the orange area in Fig. 4(b)). $PSP_1$ as a function of gradient is shown in panel (c). For positive gradients, $PSP_1$ remains nearly constant i.e. the field of the first unit cell stays localized at the interface. In contrast, for negative gradients, i.e. if the upper dispersive band approaches the energy of the defect mode, $PSP_1$ is continuously reduced, and eventually approaches 0.65. This asymmetric response with respect to the action of a gradient is due to the spectral position of the interface state, which is closer to the upper than to the lower band as observed in Fig. 4(b). It hints onto an experimental violation of chiral symmetry, most likely due to the presence of long-range coupling. But, as long as inversion symmetry is still present a localized interface state can still exist as demonstrated in Ref.[24]. Fig. 4(d) displays the simulated counterpart of panel (a). In close agreement with the experiment, we again observe the reduction in PSP and a marked increase in localization length due to the hybridization of the localized mode with the tilted dispersive bands.

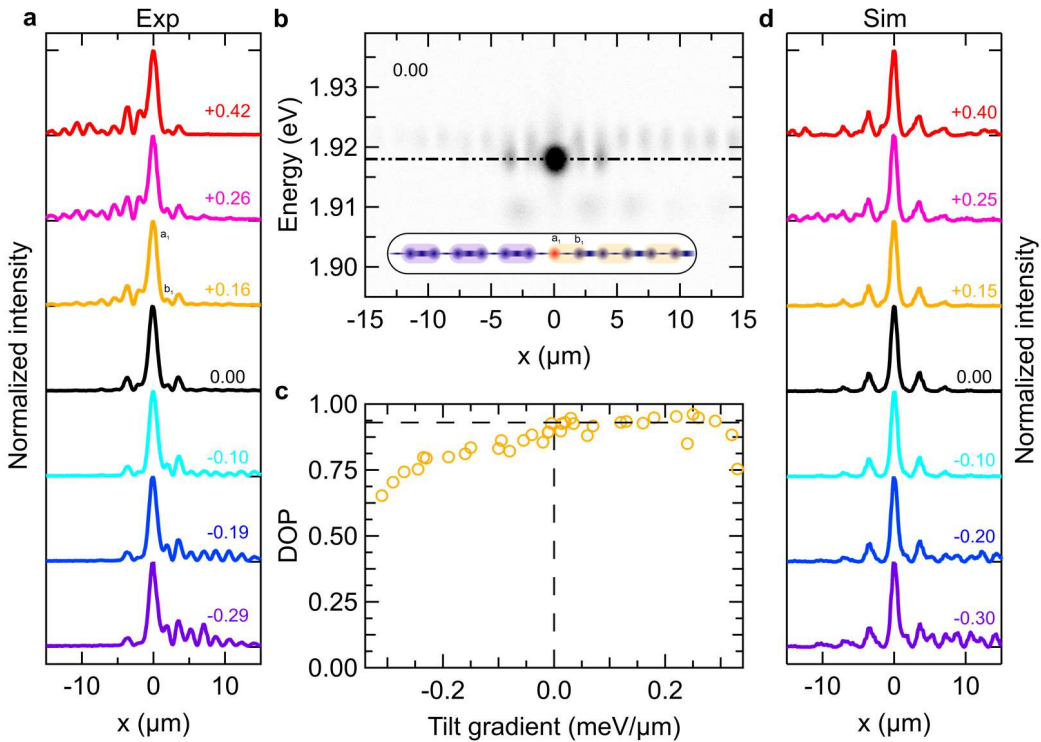

Fig. 4. **Actively tuning the topology of polaritons in a lattice a** Spatial cross-sections extracted from the measured spectra at the energy of the defect mode in the s-band for systematically sweeping of the potential gradient. Values of the gradient are indicated in meV/µm. With increasing tilt the degree of pseudo-spin polarization (PSP) drops and the localization length drastically increases towards the direction where the upper dispersive band allows tunnel coupling. **b** Spatially resolved emission spectrum for zero tilt. The dashed line indicates the spectral position at which cross-sections in (a) are extracted. The inset shows the staggered hopping and the definition of unit cells for the trivial (violet) and nontrivial (orange) case in our system. **c** PSP of the first unit cell of the nontrivial part as defined in (b) as a function of gradient. The PSP is close to 1 and is unperturbed for positive gradients. For negative gradients it gradually decreases due to energetic overlap and tunnel coupling of the localized s-band defect mode with the upper dispersive s-band. **d** Simulated cross-sections extracted at the energy of the defect mode in the s-band.

## Zak phase determination

Finally, we verify that the tilt applied to our lattice can be directly exploited to measure the topological invariants of our system, which are the Zak-phases of the two domains concatenated at the domain boundary. This experiment is conducted at a gradient of $\mu = 0.18\,meV/\mu m$. Fig. 5 displays the spatially and spectrally resolved simulation (a) and measurement (b) of our system under these conditions.

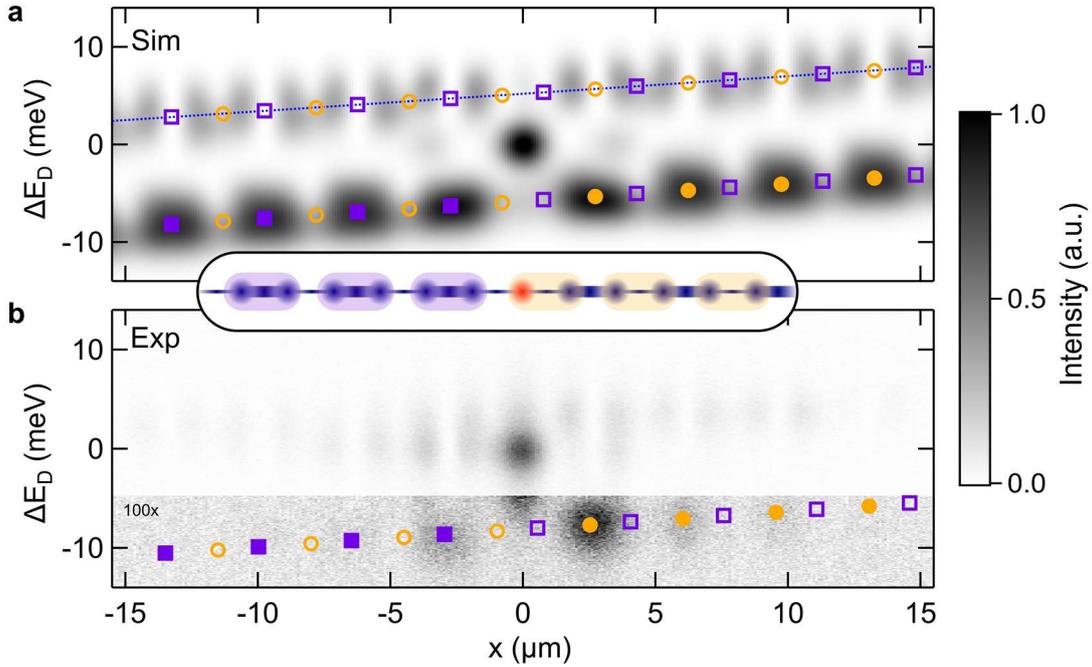

Fig. 5. **Zak phase determination.** Spatially resolved emission spectrum of the sample for a gradient of µ=0.18 meV/µm (dashed blue line). **a** Simulation of the density of states within the tight-binding approximation assuming a realistic linewidth and spatial mode profile. **b** Experimental photoluminescence spectra. The violet squares and orange circles depict Wannier-Stark ladders given by the analytical expression (see equations (7,8)). For better visibility the data below -5 meV was multiplied by a factor of 100. Inset: schematic depiction of the SSH lattice with trivial (violet) and nontrivial (orange) unit cells.

The upper panel shows the results of simulations within the single band, nearest neighbor, tight-binding approximation convoluted with a realistic linewidth and spatial mode profile. We observe, that with the applied gradient, the infinitely extended Bloch modes of the lattice transform into spatially localized Wannier-Stark states for both, the binding and antibinding s-band.

A Wannier-Stark ladder exhibits an equidistant energy spectrum[25,26], which still carries information about the topological properties of the lattice with zero gradient. As shown recently for atomic systems[27], the geometric topological phase can be extracted from the energy spectra of these Wannier-Stark states. The energies of the m-th Wannier-Stark states $\lambda_m^\pm$ are given as a function of the gradient $\mu$ and the Zak-phase $\gamma_{Zak}^\pm$ for the lower "+" and upper "-" bands as

$$\lambda_m^\pm = \frac{1}{2\pi}\int_{-\pi}^{\pi}\Omega^\pm(k)dk + \mu D \cdot \left(m + \frac{\gamma_{Zak}^\pm}{2\pi}\right), \tag{7}$$

where the periodic lattice has the dispersion relation $\Omega^\pm(k)$. Being derived first within the tight-binding model in the limit of small gradient[26], the relation (7) is also valid beyond the tight-binding approach provided that Zener tunneling is negligible. In Eq.(7) the index $m$ also labels the site, where the Wannier-Stark state is localized. Hence, it also demonstrates that the Zak phase is equivalent to a spatial shift resulting in an effective polarization of a nontrivial lattice[28]. Consequently, intensity maxima of Wannier-Stark states are found at positions

$$X_m^\pm = D \cdot \left(m + \gamma_{Zak}^\pm/2\pi\right) + X_0 , \tag{8}$$

with a constant $X_0 = D_v/2$, if the coordinate origin is chosen at the first site of the nontrivial domain. The application of this formalism to the spatially resolved energy spectrum in the tilted s-band of our lattice structure allows us to directly measure the Zak phase difference between the topologically trivial and nontrivial domains from the simulation shown in Fig. 5(a). We calculated the spectrum of the Wannier-Stark states for a fixed gradient, where the dispersion relation $\Omega^\pm(k)$ was extracted from the momentum-resolved measurements of the system without gradient (see Fig. 2(a)). For our definition of unit cells (see inset of Fig. 5) the lattice on the right hand side from the defect is topologically nontrivial with the Zak phases of $\gamma_{Zak}^\pm = \pm\pi$. The violet squares and orange circles in panel (a) depict Wannier-Stark ladders given by the analytical expression from equations (7,8), calculated for the lower ("+") and upper ("-") bands, respectively. The filled orange circles match the maxima in the topologically nontrivial part of the sample, whereas the filled violet squares match in the trivial domain. This yields a Zak phase difference of $\Delta_{Zak} = \pi$ corresponding to a shift of half of a unit cell between the topological nontrivial and trivial lattice phase for our simulation.

In Fig 5.(b) the corresponding experimental results are shown. While, in the topologically nontrivial domain the analytical results from equations (7,8) perfectly match the experiment, in the trivial domain we observe a slight offset. Formally, this can be accounted for by a measured Zak phase for the trivial domain of $\gamma_{Trivial} = -0.13\pi$ for the lower s-band. We justify this slight offset by sample imperfections which can break local inversion and chiral symmetries in the vicinity of the defect. We obtain a Zak phase difference of $\Delta_{Zak} = (1.13 \pm 0.11)\pi$, which is close to the simulated value of $\Delta_{Zak} = \pi$ within error margins.

**Conclusions**

Our demonstration of tunable polaritons in a topological lattice is of high significance: The large spectral tunability of TMD polaritons at room temperature can directly be exploited in topologically nontrivial devices of enhanced functionality. The topologically nontrivial nature of our tunable defect is of interest for the implementation of a new class of lasers, bose condensates and non-linear light-matter interfaces harnessing geometrical protection of optical modes. Furthermore, electrical injection in similar devices should become feasible with the current state of technology[29]. Last but not least, the feasibility to induce additional potentials and spatial potential gradients will become of high relevance for further studies in particular in the non-linear regime, where linear gradients of cavities are the optical counterpart of applied electrical or gravitational fields. As we furthermore demonstrate, this novel degree of freedom provides a powerful tool to directly measure and analyze topological invariants. Here, we utilized the afore mentioned tunability of our system to experimentally extract the Zak phase difference of $\Delta_{Zak} = (1.13 \pm 0.11)\,\pi$.

## Methods

### Sample

To facilitate a micrometric approach of the two DBR mirrors (<5 µm), a mesa with dimensions 100 µm x 100 µm, and 100 µm height is cut into the glass substrate of the bottom mirror before evaporating the bottom DBR (see Supplementary Note 1). The spherical cap photonic traps of the top mirror are sculpted via FIB in a glass carrier. These traps present a diameter of 5 µm and an approximate depth of ~155 nm. Details on the studied photonic lattices can be found in Supplementary Note 1. Each DBR is formed via dielectric sputtering. It is worth noting that the sputtering process in the DBR formation does not yield a planarization, or a significant roughening of the FIB structure in the top mirror.

The WS$_2$ monolayer (obtained from a CVD-grown bulk crystal) is transferred onto the bottom DBR via the dry-gel stamping method. The optical properties of the WS$_2$ monolayer are studied in Supplementary Note 1. The thickness of each SiO$_2$ [TiO$_2$] $\lambda/4$-layer is optimized for a 620 nm cavity resonance. As the bottom DBR is terminated on a low index material of $\lambda/(4n)$ thickness (SiO$_2$), the monolayer will permanently remain in the optical field antinode, and thus optimally coupled to the longitudinal optical cavity modes.

### Experimental setup

The laser source is a continuous wave, green laser; its pump power is controlled with adjustable optical density filters in the excitation path. The laser is tightly focused in the top mirror of the cavity with an Mitutoyo objective (×50, 0.65 NA). This allows to form a Gaussian spot size with a FWHM diameter of ~3.6 µm. The spectrometer [CCD] used for these experiments is an Andor Shamrock 500i [Andor iKonM 934]. Further details on the experimental setup for the reconstruction of the real- and momentum-space distribution are given in the Supplementary Note 1.

### Acknowledgement

The authors gratefully acknowledge funding by the State of Lower Saxony. We acknowledge funding by the Lower Saxony ministry for science and education (MWK) within the Landesgraduiertenkolleg "DyNano". Funding by the German Research Society (DFG) within the project SCHN1376 11.1 and 14.2 is acknowledged. ST acknowledges support from DOE-SC0020653 and NSF DMR 2111812 and GOALI CMMI-2129412. FE is supported by the Federal Ministry of Education and Science of Germany under Grant ID 13XP5053A. M.E. acknowledges funding by the University of Oldenburg through a Carl von Ossietzky Young Researchers Fellowship.

# Supplementary Information

# Topologically tunable polaritons based on two-dimensional crystals in a photonic lattice

L. Lackner[1], O.A. Egorov[2], A. Ernzerhof[1], C. Bennenhei[1], V.N. Mitryakhin[1], G. Leibeling[3], F. Eilenberger[3,4], S. Tongay[5], U. Peschel[2], M. Esmann[1] and C. Schneider[1,†]

[1]*Institut für Physik, Fakultät V, Carl von Ossietzky Universität Oldenburg, 26129 Oldenburg, Germany.*

[2]*Institute of Condensed Matter Theory and Solid State Optics, Friedrich Schiller University, Jena, Germany*

[3]*Fraunhofer-Institute for Applied Optics and Precision Engineering IOF, 07745 Jena, Germany.*

[4]*Institute of Applied Physics, Abbe Center of Photonics, Friedrich Schiller University, 07745 Jena, Germany*

[5]*School for Engineering of Matter, Transport, and Energy, Arizona State University, Tempe, Arizona 85287, USA*

[†]*Corresponding author. Email: Christian.Schneider@uni-oldenburg.de*

### *Supplementary Note 1. Sample, experimental open cavity setup and PL imaging system*

Tungsten disulfide ($WS_2$) vdW crystals were grown using a two-step flux technique. As received tungsten (5N purity Alfa Aesar) and sulfur (5N purity Sigma Aldrich) were further purified to 6N purity or higher using electrocatalytic reaction and/or sublimation technique. These high purity powders (300 mesh) were mixed till a uniform mixture was formed using a 14-days process using an automated powder tumbler process. The uniform mixture contained a stoichiometric ratio of W:S (1:2 molar) with access sulfur to ensure a close to 1:2 ratio. The mixture was sealed under 1E-7 Torr pressure in a 2 mm thick quartz ampoule that measured 19 cm in length and baked at high temperatures (1000 °C ramp rate 20 °C / hour, natural cooling) till polycrystalline ~200-100 mesh binary $WS_2$ vdW powders were created. The usual process required a 1 to 3 weeks annealing process. After the first step, these powders were removed from the ampoule under argon backfilled glovebox and further ground using the tumbler process to create uniform size density vdW $WS_2$ powders.

In the 2$^{nd}$ step, these powders were sealed in the ampoule without any transport agent under 1E-7 Torr or better pressure with 2 mg/1 g extra sulfur concentration. The access sulfur was essential to obtaining defect-free $WS_2$ crystals without bound exciton peaks at low temperatures. The sealed ampoule was ramped very slowly to 1210 °C at 20 °C/hour and kept at this temperature for 3 days. One end of the ampoule was gradually dropped to 1170 °C (temperature differential of 40 °C) over 2 days. The entire process (2$^{nd}$ segment) took 5 weeks to yield a few mm-sized self-flux grown crystals without any transport agent aide to limit contamination arising from $I_2$, $Br_2$, or other transport agents. In the entire process, the purity of the powders, formation of high mesh number vdW powders as precursors, careful cold end temperature drop, and the duration of the crystal were essential to achieving excellent excitonic grade $WS_2$ crystals.

The atomically thin $WS_2$ layer is mechanically exfoliated from the grown bulk crystal and then transferred onto the bottom DBR via the dry-gel stamping method[1]. After the transfer of the $WS_2$ monolayer the sample is annealed at 100 °C on a hot plate for 1 min.

To facilitate a micrometric approach of the two distributed Bragg reflector (DBR) mirrors (<3 µm), we have cut away ~100 µm of the glass substrate sparing a mesa of dimensions 100 µm

x 100 µm, and 100 µm height, before evaporating the bottom DBR[2] (see Supplementary Figure 1(a) where the limits of the mesa surface are visible). The bottom [top] DBR consist of 10 [8] pairs $SiO_2/TiO_2$. The thickness of each $SiO_2$ [$TiO_2$] $\lambda/4$-layer is optimized for a 620 nm cavity resonance. The cavity is terminated on $SiO_2$ (lower refractive index) to ensure the maximum coupling of the $WS_2$ monolayer to the photonic cavity field.

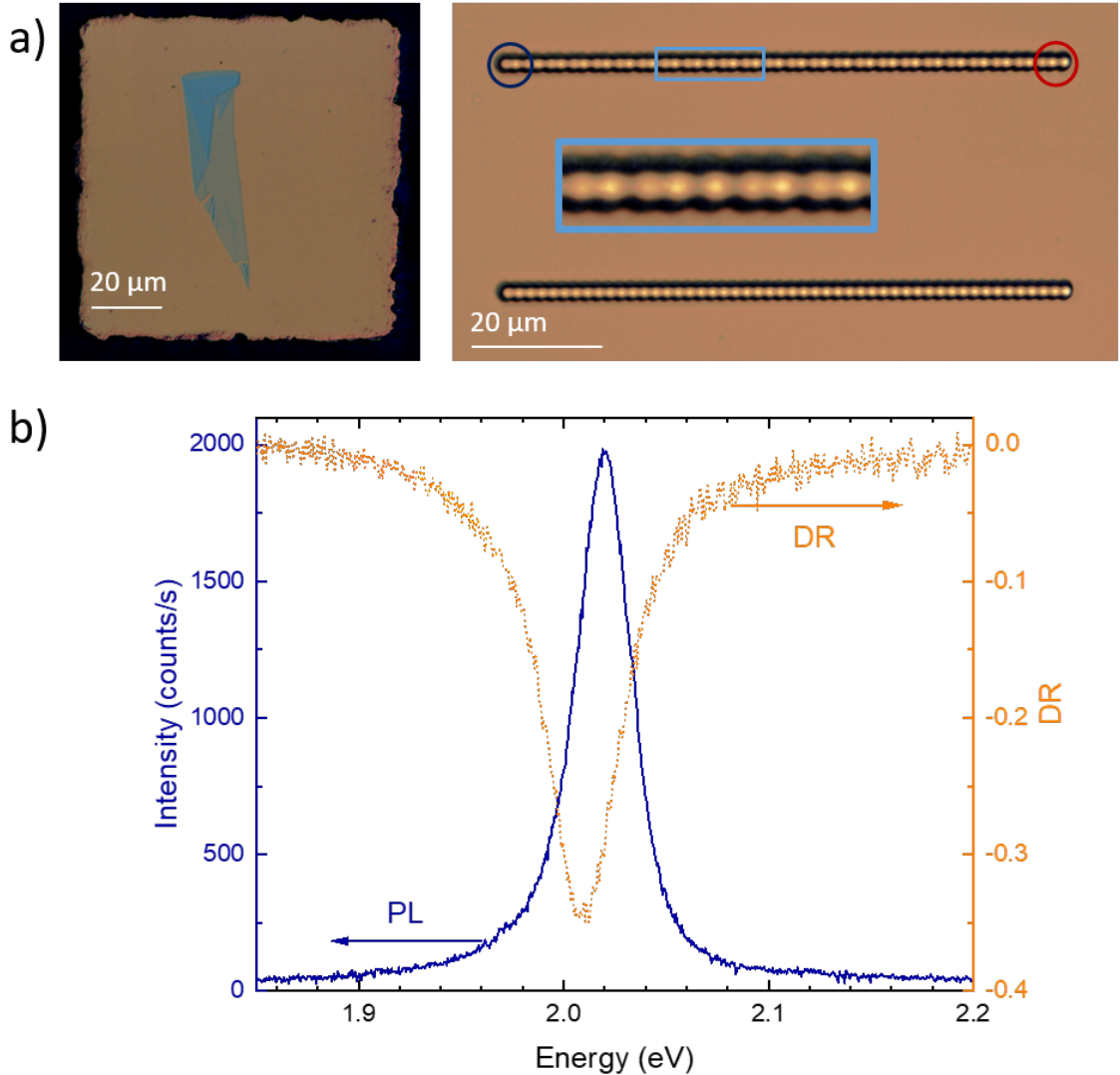

**Supplementary Figure 1. Detail on the DBR mirrors, PL and DR of the $WS_2$ monolayer.** a) Microscope images of the two DBR mirrors. Left: $WS_2$ monolayer (light blue) transferred onto the bottom DBR. The edges of the mesa are clearly visible. The monolayer has a size on the order of 10µm by 30µm without wrinkles or other visible defects. Right: Detail of the photonic potentials implemented in the top DBR. All presented chains have a single site diameter diameter $d = 3.00$ µm. On the top a 1D chain with staggered site center-to-center distances $D_v = 1.95$ µm and $D_w = 1.56$ µm (SSH chain) is shown. The two end of chain boundaries are marked with circles. The blue circle identifies a boundary with strongly overlapping sites (large hopping) and the red circle the weakly overlapping sites (weak hopping). The box in the center (light blue) is zoomed in on the central domain boundary with the "defect" site studied in this work. On the bottom a regular 1D chain with site center-to-center distance $D_w$ is shown. The white scalebars correspond to 20µm. b) PL (blue) and DR (orange) shows spectra of the WS2 monolayer studied without top DBR. The PL spectrum was acquired using a pump power of ~10 µW and 1 s integration time. The white light reflectivity spectra were recorded using a quartz tungsten-halogen lamp (Thorlabs SLS301) and an integration time of 0.1 s. The differential reflectivity is calculated by $DR = (Sample - Reference)/Reference$, where the reference signal was acquired on the bare top mesa without any active material.

The WS$_2$ monolayer in Supplementary Figure 1a) is of dimensions of 30 µm x 10 µm. In panel b) corresponding PL (blue trace) and DR (orange trace) spectra of the monolayer are shown. The spectral position of the exciton PL[DR] is 2.019[2.010] eV. The corresponding FWHM of these peaks is 30[39] meV. On the right side Supplementary Figure 1a) shows a microscopy image of the structured top DBR mirror. The spherical cap photonic traps of the top mirror are fabricated via gallium Focused Ion Beam (FIB) in a glass substrate (SCHOTT 'D 263® T eco Thin Glass', thickness=550 µm). These traps are of varying diameter and shape. Here only one set of one-dimensional linear SSH chains is shown. The chain consists of 50 individual sites with fixed diameter $d = 3.00$ µm and depth $t \approx 155$ nm, corresponding to the cap of a sphere with a radius of curvature $r_c = (t^2 + (d/2)^2)/(2t) = 7.34$ µm. The site overlap is varied by the center-to-center distance D$_{v,w}$ with $D_v = 1.95$ µm and $D_w = 1.56$ µm. The two natural domain boundaries of the chain are designed to end on i) a small overlap D$_v$ (weak hopping) and ii) a big overlap D$_w$ (strong hopping). Additionally a third domain boundary is generated inside of the chain (Position 19) combining two lattices with Zak-phase 0 and $\pi$, respectively.

The air-gap open cavity is realized by attaching both DBR mirrors to sets of nano-positioners (see Supplementary Figure 2). The bottom DBR is mounted on a stack of five nano-positioners. From bottom to top: XY- (12 mm travel range, model ecsxy5050alnumrt) and Z-positioner set (8 mm travel range, ecsz5050alnumrt), followed by tilt-positioners forming a goniometer (see labels θ and φ in panel (b), models ecgt5050alnumrt and ecgp5050alnumrt) and a rotator (rot, providing 360 deg rotation, model ecr3030alnumrt). The top DBR is mounted on a identical set of XY- and Z-positioners as for the bottom DBR. Tip and tilt are aligned to a desired fixed position when mounting the cantilever (see Supplementary Figure 2) carrying the top DBR.

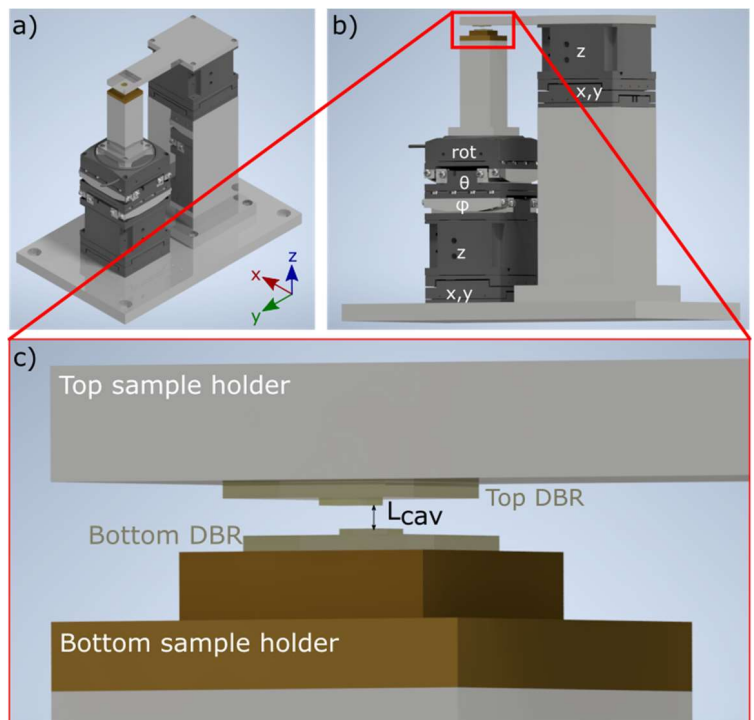

**Supplementary Figure 2. CAD-drawing of the air-gap open cavity system.** (a,b) Overview of the air-gap open cavity assemble in different perspectives. The red box in panel (b) indicates the area where the sample is loaded in the microcavity. The different labels in panel (b) indicate all the degrees of freedom controlled by the motors in order to align the open cavity system. (c) Close-up of the sample area. The two DBR mirrors are pasted to individual sample holders (the monolayer is deposited on top of the bottom DBR). The bottom/top sample holder is made out of bronze/steel.

Supplementary Figure 3 shows a sketch of the lens configuration to collect angle- (momentum-space) and real space-resolved (real-space) emission from the sample in the experiments. For the momentum-resolved measurements a Fourier imaging configuration is used. Lens L$_K$ collects the angle-dependent information of the back-focal plane of the microscope objective L$_{Obj}$ (Mitutoyo Plan Apo NIR HR 0.65 NA, 50x). To collect real-space resolved PL an extra lens L$_R$ is inserted in the beam path. The respective images are projected to the spectrometer focal plane (slit) by L$_S$.

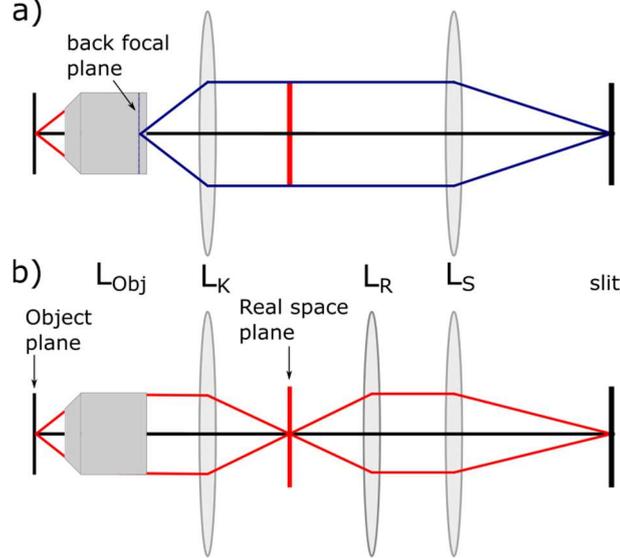

**Supplementary Figure 3. Optical system scheme for real- and momentum-space imaging.** a) A Fourier imaging configuration is used to create an image in the focal plane of the spectrometer system (slit). The Fourier lens (L$_K$, f=300 mm) collects the angle-resolved information in the back focal plane of the microscope objective (L$_{obj}$, ×50, 0.65 NA) which is imaged to the spectrometer focal plane by the collection lens (L$_S$, f= 400mm). b) To collect the real-space distribution PL an extra lens (L$_R$, f=200 mm) is inserted in the beam path.

## *Supplementary Note 2. Description of the coupled oscillator model*

To describe the upper and lower polariton resonances, we employ a standard two-coupled-oscillators model:

$$\begin{bmatrix} E_{ex} & \hbar\Omega_R \\ \hbar\Omega_R & E_{cav} \end{bmatrix} \begin{bmatrix} X \\ C \end{bmatrix} = E \begin{bmatrix} X \\ C \end{bmatrix}, \quad (S.2.1)$$

where $E_{ex}$ and $E_{cav}$ denote the energies of the exciton and cavity modes, respectively, and $2\hbar\Omega_R$ is the normal mode splitting. For the lower polariton branch the Hopfield coefficients $X$ and $C$ are given by:

$$|X|^2 = \frac{1}{2}\left(1 + \frac{E_{cav} - E_{exc}}{\sqrt{(E_{cav} - E_{exc})^2 + 4(\hbar\Omega_R)^2}}\right) \quad (S.2.2)$$

$$|C|^2 = 1 - |X|^2 \quad (S.2.3)$$

Their squared amplitudes $|X|^2$ and $|C|^2$ quantify the exciton and cavity photon fractions. The eigenenergies of the upper and lower polariton branches are obtained by solving the eigenvalue problem:

$$E_{UP,LP}(k_{||}) = \frac{1}{2}\left(E_{ex} + E_{cav} \pm \sqrt{4(\hbar\Omega_R)^2 + (E_{cav} - E_{exc})^2}\right) \quad (S.2.4)$$

### Supplementary Note 3. Bloch Modes within the mean-field model with effective potential

In order to determine the energy-momentum band structure of polaritons in the SSH lattice, we calculate Bloch modes in the mean-field approximation, where the geometry of the chain is represented by an effective potential in two transverse (x and y) directions. The mean-field approach is valid in the vicinity of a longitudinal resonance of the cavity and requires that the respective longitudinal mode profile between the mirrors (z-direction) is fixed. Then, in the first approximation of the perturbation theory, it is possible to reduce the three-dimensional problem to respective two-dimensional one (x- and y-) by separating the longitudinal mode profile (z-direction).

Applying this mean-field approach we solve the following eigenvalue problem for the energy $E(k_{||})$ of the Bloch mode with the Bloch vector $\boldsymbol{k} = k_{||}\vec{e}_x$:

$$E(k_{||})\begin{Bmatrix} p_b(\mathbf{r}, k_{||}) \\ e_b(\mathbf{r}, k_{||}) \end{Bmatrix} = \hat{L}(k_{||})\begin{Bmatrix} p_b(\mathbf{r}, k_{||}) \\ e_b(\mathbf{r}, k_{||}) \end{Bmatrix}, \quad (S.3.1)$$

where the functions $p_b(\mathbf{r}, k_{||})$ and $e_b(\mathbf{r}, k_{||})$ describe the amplitude distributions of the photonic and excitonic components of the Bloch modes in real space defined by the resonator $\mathbf{r} = \{x, y\}$. The main matrix in Eq. (S5.1), describing the single-particle coupled states of excitons and photons, is given by the expression

$$\hat{L}(k_{||}) = \begin{pmatrix} \omega_C^0 + V(\boldsymbol{r}) - \frac{\hbar}{2m_C}(\vec{\nabla}_\perp + ik_{||}\vec{e}_x)^2 & \Omega_R \\ \Omega_R & \omega_E^0 - \frac{\hbar}{2m_E}(\vec{\nabla}_\perp + ik_{||}\vec{e}_x)^2 \end{pmatrix} \quad (S.3.2)$$

In the model above, the quantities $\omega_C^0$ and $\omega_E^0$ represent the energies of bare photons and excitons, respectively. The photon-exciton detuning used for the calculations is $\hbar\omega_C^0 - \hbar\omega_E^0 = -293\ meV$. The photon-exciton coupling strength is given by the parameter $\Omega_R$ which defines the Rabi splitting $2\hbar\Omega_R = 20 meV$ between coupled excitons within TMD and photons of the cavity mode. The kinetic energy of polaritons is characterized by the effective mass $m_C \approx 5.3 \cdot 10^{-6} m_e$ ($m_e$ free electron mass) which defines transport properties of the intracavity photons. The effective exciton mass is $m_E = 10^5 m_C$.

The periodic array (with the period $D = D_w + D_v = 3.51\ \mu m$) is modeled by the two-dimensional potential $V(r)$ and consists of spatially-overlapping polaritonic traps. A single cell of the SSH lattice consists of two spherically-shaped potential profiles defined as the real part of the function:

$$V(x,y) = \frac{V_0}{\left(\sqrt{1-\left(\frac{d}{2r_c}\right)^2}-1\right)} \left(\begin{array}{l} \left(\sqrt{1-\frac{(x+D_{v,w}/2)^2+(y)^2}{r_c^2}}-1\right), \ for\ -\frac{D}{2}<x\leq 0 \\ \left(\sqrt{1-\frac{(x-D_{v,w}/2)^2+(y)^2}{r_c^2}}-1\right), \ for\ \ 0<x\leq\frac{D}{2} \end{array}\right) \quad (S.3.3)$$

The intra-cell distance $D_{v,w}$ is equal to either $D_v$=1.95 µm or $D_w$=1.56 µm for topologically nontrivial or topologically trivial cases. The potential depth is $V_0 = 140\ meV$ and the spatial size of the traps is $d = 3\mu m$. The parameter $r_c$ describes the radius of the indentations in the upper mirror as introduced in Supplementary Note 1.

**Supplementary References**

1. Castellanos-Gomez, A. *et al.* Deterministic transfer of two-dimensional materials by all-dry viscoelastic stamping. *2D Mater.* **1**, 301–306 (2014).

2. Knopf, H. *et al.* Integration of atomically thin layers of transition metal dichalcogenides into high-Q, monolithic Bragg-cavities: an experimental platform for the enhancement of the optical interaction in 2D-materials. *Opt. Mater. Express* **9**, 598 (2019).